\documentstyle[aps,prd]{revtex}
\def\ga{\mathrel{\raise.3ex\hbox{$>$\kern-.75em\lower1ex\hbox{$\sim$}}}}
\def\la{\mathrel{\raise.3ex\hbox{$<$\kern-.75em\lower1ex\hbox{$\sim$}}}}
\begin{document}
%%%%%%%%%%%%%%%%
\title{Update of the direct detection of dark matter 
	and the role of the nuclear spin} 
\author{V.A.~Bednyakov}
\address{Laboratory of Nuclear Problems,
	 Joint Institute for Nuclear Research, \\ 
	 141980 Dubna, Russia; E-mail: bedny@nusun.jinr.ru \\[1mm] 
	{\normalsize\rm and} \\[-3mm] } 
\author{H.V.~Klapdor-Kleingrothaus }
\address{
        Max-Planck-Institut f\"{u}r Kernphysik, \protect\\
        Postfach 103980, D-69029, Heidelberg, Germany; \\
	E-mail: klapdor@gustav.mpi-hd.mpg.de}
\maketitle

\begin{abstract}
	We update our exploration of the MSSM parameter space at the weak 
	scale where new accelerator and cosmological constraints are respected.
	The dependence of WIMP-nucleon cross sections on parameters of 
	the MSSM, uncertainties of the nucleon structure and other 
	theoretical assumptions like universality and co-annihilation 
	are considered. 
	In particular, we find that the coannihilation
	does not have a significant effect in our analysis in certain 
        regions which are allowed even with coannihilation.
	The new cosmological constraint on the relic neutralino density 
	used in the form $0.1 < \Omega_\chi h^2_0<0.3$
	does also not significantly affect the regions of allowed 
	neutralino-nucleon cross sections.
	We notice that for nuclear targets with spin the spin-dependent 
        interaction may  
	determine the lower bound for the direct detection rate
	when the cross section of the scalar interaction 
	drops below about $10^{-12}\,$pb.
\end{abstract}
\vspace{0.5cm} %\pacs{11.30.Pb,12.60.Jv,95.35.+d}

%%%%%%%%%%%%%%%%%%%%%%
\section{Introduction}  
%%%%%%%%%%%%%%%%%%%%%%
	It is well known (see, for example, 
\cite{h0007202})
	that the Minimal Supersymmetric Standard Model (MSSM) being 
	the most promising extension of the Standard Model offers a 
	solution of the hierarchy problem, 
	possesses gauge coupling unification and naturally proposes 
	a Dark Matter candidate~--- the 
	lightest supersymmetric particle (LSP). 
	In the framework of the low-energy supersymmetry (SUSY),
	when SUSY breaking masses lie below a few TeV, 
	sparticles will be copiously produced (and detected) at
	future colliders like the Large Hadron Collider (LHC) at CERN.
	On the other hand, there are several on-going
	and future projects searching for the 
	LSP as a Dark Matter particle.
	One of them even claims a positive signal
\cite{DAMA}, although the situation remains rather contradictory
\cite{CDMS2000}. 
	The present experimental upper limit
	on the spin-independent part of the 
	elastic scattering of the LSP on a nucleon is around $10^{-5}$~pb
	for $50~{\rm GeV} \la m_\chi \la 100$~GeV. 
	In any case, it seems very plausible that both, SUSY collider 
	signals and LSP Dark Matter will be found in future.
	Such dark matter searches offer interesting 
	prospects for beating accelerators in the discovery of 
	SUSY, particularly during	
	the coming years before the LHC enters operation
\cite{h0001005}.
	In this situation naturally arised the question 
	of how small the event rate of the 
	LSP direct detection can be, provided the LSP is a cold dark
	matter particle.
	Searching for the answer 
	different SUSY models were considered (see for example, 
\cite{h9701301,h9908427,h0003186,h0008022,h0008336}).

	Recently exciting evidence for a flat and 
	accelerating universe was obtained 
\cite{a0004404,a0005124}.
	The position of the first acoustic peak of the angular power spectrum
 	strongly suggests a flat universe with density parameter 
	$\Omega_{0}=1$ 
	while the shape of the peak is consistent with the
	density perturbations predicted by models of inflation.
 	The density parameter  
	$\Omega_0 = \rho_0/\rho_c$ is the ratio of the current 
	mass density $\rho_0$ to  the critical density
   	$\rho_c = 1.88\cdot 10^{-29}\,h^2_0\,$g$\cdot$cm$^{-3}$,
	with $h_0$ being the dimensionless Hubble parameter.
	Data support the straight line
	$\Omega_0 = \Omega_{M} + \Omega_{\Lambda}=1$
\cite{a9912211,a0007333,a0004393,h0008145}, 
	where $\Omega_{M}$ is the matter density in the universe and 
	$\Omega_{\Lambda}$ is the contribution of the non-zero
	cosmological constant (the energy density of the vacuum).
	At the same time one determines $\Omega_{M}= 0.4 \pm 0.1$, 
	which implies $\Omega_{\Lambda} =0.85\pm 0.2$, 
	a value that has been supported from high-redshifted %$z$ 
	Supernova data 
\cite{SNa}.
	Since the baryonic matter density is small, 
	${\Omega}_{B}=0.05 \pm 0.005$, the values for matter density 
	$\Omega_{M}$ give a Cold Dark Matter (CDM) density 
	${\Omega}_{\rm CDM} \simeq 0.35 \pm 0.1$, which combined with 
	recent measurements of the Hubble parameter $h_0=0.65 \pm 0.05$,  
	result in smaller CDM relic densities 
	$\Omega_{\rm CDM} \, {h_0}^2 \simeq 0.15 \pm 0.07 $
\cite{a9912211,a0007333,a0004393,h0008145}.

	Previously we have restricted our analyses to the cosmological 
	constraint for the relic density of the LSP in the range 
	$0.025 < \Omega_\chi h^2_0<1$
\cite{DressNojiriRate,h9606261,h9608241,h9801246}, where the 
	neutralino relic density parameter   
	$\Omega_\chi = \rho_\chi/\rho_c$ and $\rho_\chi$ is 
	the relic neutralino mass density. 
	In this paper adopting the above-mentioned new cosmological data 
	and going to compare our estimation with literature
\cite{h0001005,h0008336,h0009065,h0010203}
	we constrain the allowed region for the neutralino 
	relic density in the form $0.1 < \Omega_\chi h^2_0< 0.3$.
	It is possible that there is more than one component in the cold dark
	matter, so that $\Omega_\chi < \Omega_{\rm CDM}$, and 
	therefore $\Omega_\chi < 0.1$. 
	Although, in general, a decrease of $\Omega$ is associated with 
	larger elastic scattering cross sections, the detection
	rate also must be reduced because of the corresponding reduction
	in the density of LSPs in the Galactic halo. 
	Here we neglect this possibility and assume that all the cold 
	dark matter is composed of LSPs, so that $\Omega_\chi \ge 0.1$
\cite{h0001005}.

\smallskip
	There are two main approaches to 
	evaluate the dark-matter-neutralino nucleon cross section
	and the expected event rate in a detector. 
	The basis of the first approach is the 
	minimal supergravity (mSUGRA) model 
\cite{mSUGRA}.
	This model assumes the minimal supersymmetric standard model
 	to be valid at all energy scales from $M_{\rm weak}$ up to 
	$M_{\rm GUT}\simeq 2\times 10^{16}\,$GeV. 
	The mSUGRA model arises as the low-energy 
	limit of a supergravity theory, 
	where supersymmetry is broken in the
	hidden sector of the model at an energy scale of $M\sim 10^{10}$ GeV. 
	Supersymmetry breaking is communicated to the observable sector via 
	gravitational interactions, leading to soft SUSY breaking mass terms
	of the order of the electroweak scale. 
	At the GUT scale this leads to a common mass 
	for all scalars $m_0$ and a common trilinear coupling $A_0$. 
	Motivated by the apparent unification of gauge coupling 
	constants, it is also assumed that all 
	gaugino masses are unified to
	$m_{1/2}$ at $M_{\rm GUT}$. 
	The weak scale sparticle spectrum is derived from 
	renormalization group running of the SUSY soft breaking parameters.
	Requiring radiative electroweak symmetry 
	breaking allows the determination of the superpotential 
	Higgsino mass squared $\mu^2$, and allows the expression of
	the soft SUSY breaking bilinear term $B$ in terms of 
	$\tan\beta$. 
	Thus, all sparticle masses and couplings are 
	derived in terms of the {minimal}\ parameter set 
$m_0,\ m_{1/2},\ A_0,\ \tan\beta ,\ {\rm and}\ {\rm sign}(\mu)$
\cite{mSUGRA,h9706509,h0003154}.
	From a practical point of view this  approach as much as possible 
	relies on theoretical arguments like unification, 
	naturalness etc, aiming to maximally reduce the set of 
	free parameters and obtain maximally restricted predictions.
	In this approach 
\cite{h0001005} the quantum stability of the gauge hierarchy suggests that 
	sparticles weigh less than about 1~TeV
\cite{hierarchy}, which is also the range favored for a cold
	dark matter particle, and there are indeed generic domains 
	of the MSSM parameter space in which the relic LSP density 
	falls within the range favored by astrophysics and cosmology. 
	The unsuccessful laboratory searches for sparticles 
	impose non-trivial constraints on the MSSM parameter space, 
	suggesting that the LSP neutralino is mainly a $U(1)$
	gaugino (Bino)
\cite{EFGOS}.
	In the MSSM the lightest neutralino 
	$\chi \equiv \tilde\chi_1^0$
	is a mixture of four superpartners of gauge and Higgs bosons
	(Bino, Wino and two Higgsinos):
\begin{equation}
\label{LSPmixture}
\chi =   N_{11}\widetilde B^0  +N_{12}\widetilde W^0
	+N_{13}\widetilde H_1^0+N_{14}\widetilde H_2^0.
\end{equation}
	It is commonly accepted that $\chi$ is mostly gaugino-like if 
	$P\equiv N^2_{11}+N^2_{12}>0.9$ and  
	Higgsino-like if $P < 0.1$, or mixed otherwise.
	
	It is due to the Bino-likeness of the relic neutralinos
	that the calculated neutralino-nucleon 
	cross sections appeared very small and one usually 
	arrived at the conclusion that it was hardly possible to reach the 
	mSUGRA space by means of direct and indirect searches 
	for dark matter particles
\cite{h0001005,h9701301,h0008022,DressNojiriRate,h9706509,KaKoRozWells,h9509260,h9904497}.
	The other side of this conclusion is also well known:
	when $|\mu|$ decreases, the Higgsino components $N_{13}$ and $N_{14}$
	of $\chi$ increase ($P$ decreases) and as a result the 
	spin-independent cross section increases.
	So Higgsino-like (and mixed) neutralino on the other hand 
	increases the prospect for its detection as a dark matter particle
\cite{h0003186,h0010203,h0004043,h0006266}.	
	Therefore it seems a crucial question here, to what extent the 
	neutralino is mostly gaugino-like, Higgsino-like, or mixed. 

	A way to look for any possibility of 
	higher cross sections and higher expected rates of detection is 
        to investigate alternate models.  
	The basis of it is a departure from the stringent mSUGRA
	by means of a relaxation of some unification and other theoretical 
	assumptions aiming to obtain as general predictions for 
	the expected detection rate as possible.
	A remarkable shift from mSUGRA to more relaxed models was made by 
\cite{h9801246,h0007202,h0008336}.
	It mostly included relaxation of unification of soft scalar mass 
	parameters (so called nonuniversal soft symmetry breaking)
	as well as gaugino mass non-universality. 
	The large $\tan\beta$ regime was also considered as a source of 
	higher cross sections.
	Indeed in canonical mSUGRA it was pointed out
\cite{Bottino,arna,h0001019} that the large $\tan\beta$ regime   
	allows regions where $\sigma_{\chi\,p}\approx 10^{-6}\,$pb.
	Besides,  with non-universal soft scalar masses,
	it was also found that $\sigma_{\chi\,p}\approx 10^{-6}\,$pb 
	for small values of $\tan\beta$.
	In particular, this was obtained for 
	$\tan\beta\ga 25$ ($\tan\beta\ga 4$) 
	working with universal (non-universal) soft terms in 
\cite{h0001019}.
	These analyses were performed assuming (non-)universality 
	of the soft breaking terms at the unification scale, 
	$M_{\rm GUT}\approx 10^{16}$ GeV, which
	can be obtained within superstring theories 
\cite{Witten,Rigolin} and heterotic M-theory 
\cite{Witten,Banks}.

	Completely new possibilities have also been discussed. 
	For example, it is found that in supersymmetry models muli-TeV 
        scalar masses can exist consistent with naturalness on a certain 
        branch of the radiative breaking of the electroweak symmetry 
\cite{h9710473}. 
        A similar phenomenon appears in the so-called focus point 
        supersymmetry models where one also avoids naturalness 
        constraints with multi-TeV scalars 
\cite{h0004043,h9908309,a0008115}, 
	and in models with moving intermediate unification scale
\cite{h0006266}.

	It was noticed that the assumptions concerning universality 
	of the scalar masses $m_i(M_{\rm GUT}) \equiv m_0$,
	and the trilinear scalar couplings
        $A^{l,u,d}(M_{\rm GUT}) %= A^{d}(M_{\rm GUT}) = A^{u}(M_{\rm GUT}) 
	\equiv A_0$,
	are not very solid, at least from phenomenological point of view,
	since, universality might occur 
	at a scale higher than $M_{\rm GUT}\sim 10^{16}$~GeV 
\cite{com}, or according to string models 
	at a scale $M_I$ smaller than $M_{\rm GUT}\sim 10^{16}\,$GeV
\cite{h0006266,h0005260}. 
	It was realized that the string scale may be anywhere between
	the weak scale and the Planck scale. 
	For instance D-brane configurations
	allow these possibilities in type I strings 
\cite{Lykken,Dimopoulos,typeITeV,typeIinter}.
	Similar results can also be obtained in type II strings 
\cite{typeII} and weakly and strongly coupled heterotic strings 
\cite{stronghete,weakandstronghete}.
	Moreover the $M_I$ might be anywhere between the weak
	scale and the Planck scale
\cite{h0006266}, with significant consequences  for the size of
	the neutralino-nucleon cross section.  
	The case of non-universal gaugino masses was analyzed in 
\cite{h9908427,h0003186,griest,h0001019}
	and with respect to direct detection of the superlight 
	dark matter neutralinos in
\cite{h9608241}. 
	Schemes with CP-violating phases one can find in 
\cite{CPphases}. 

   	Therefore due to the large uncertainties involved 
	in the choice of the scale $M_I$
	and going to obtain as much as general predictions 
	it appeared  more convenient to work within 
	a phenomenological SUSY model 
	whose parameters are defined directly at the electroweak scale
	as for example in  
\cite{h9908427,h0008022,h9510252,h0005171,h9802344,h0002126}
	and which is denoted as an effective scheme of MSSM (effMSSM) in 
\cite{h0010203}. 

	Obviously, this way much larger expected event
	rates were obtained and optimistic conclusions 
	concerning the possibility to 
	constrain significantly the 
	SUSY parameter space with dark matter experiments
	were drawn
\cite{h0008022,h0010203,h0004043,h0006266,h9510252,EF}.

	In our previous calculations in effMSSM
\cite{h9908427,h9606261,h9608241,h9401262} we have adopted an (effective) 
	scheme (with non-universal scalar masses and with non-universal 
	gaugino soft masses) which has supplied us with large relatively  
	direct detection rates of dark matter neutralinos, 
	practically independently of what is the neutralino composition.
	In most of the MSSM parameter space we, in agreement
	with others, have obtained at the detectable level  
	mostly gaugino-like neutralinos, but always existed small 
	Higgsino admixtures (at a level less than 1-5\%) which
	managed to produce large enough cross sections and rates.

\smallskip
	In 1994 we claimed that nuclear spin is not important 
	for detection of dark matter particles, 
	provided the detection sensitivity 
	does not exceed 0.01 events$/$day$/$kg, 
	which was considered that time as unreachable
\cite{h9401262}.
	Now the situation has changed and we would like to notice 
	that for targets with spin-non-zero nuclei
	it might be the spin-dependent interaction that 
	determines the lower bound for the direct detection rate
	when the cross section of the scalar interaction, which is 
	usually assumed to be the dominant part, 
	drops below $10^{-12\div13}\,$pb
\cite{h9908427}.

	New updated parameters of the nucleon structure
	involved in the evaluation of the elastic neutralino nucleon
	scattering have become available 
\cite{h0001005} and one expects that they will affect
	the cross sections of neutralino nucleon scattering.
	At least significant cancellations may occur for 
	some values of $\tan\beta$ for scalar- and spin-dependent 
	cross sections (at least for $\tan\beta< 10$ 	
\cite{h0001005}).

	The above considerations stimulated us
	to perform a recalculation of our previous analysis.
 
%%%%%%%%%%%%%%%%%%
\section{Approach}  
%%%%%%%%%%%%%%%%%%
        A dark matter event is elastic scattering 
	of a relic neutralino from a target nucleus producing a nuclear 
	recoil which can be detected by a suitable detector
\cite{JKG}.
	The differential event rate in respect to the recoil 
	energy is the subject of experimental measurements.
	The rate depends on the distribution of
        the relic neutralinos in the solar vicinity and
        the cross section of neutralino-nucleus elastic scattering.
	In our analysis we use the total event rate $R$ 
	which is integrated over recoil energies and useful 
	for searching for domains with extreme rates.
	We follow our papers
\cite{h9606261,h9608241}, where one can find all relevant formulas
	and astrophysical parameters.
	We consider only a simple spherically symmetric isothermal 
	distribution and do not go into details of any possible
	uncertainties (and/or modulation effects) 
	of the Galactic halo WIMP distribution 
\cite{a9808320,h9803295,a0008156,a0008318,a0009467,h0006183,h0010151}.

	To calculate the event rate we use for the relic neutralino 
	mass density and for the escape neutralino velocity commonly 
	accepted values 0.3~GeV$/$cm$^3$ and 600 km$/$s, respectively.
	Their experimental variations can slightly change $R$ but 
	leave the dependence of $R$ on the MSSM parameters unaffected.
	To compare our results with other calculations and sensitivities of 
	different dark matter experiments we calculate also the total cross 
	section for relic neutralino elastic scattering on the nucleon.
	The scalar (spin-independent) part of the elastic 
	neutralino-proton(neutron) cross section at 
	zero momentum transfer $q=0$ is 
$$
\sigma^{p,n}_{{\rm SI}}(0) 
	= 4 \frac{\mu^2}{\pi} \left[ {\cal C}_{p,n} %%%%% f_{p,n} 
		\right]^{2},
\qquad {\rm where} \qquad %\frac{
{\cal C}_{p,n} 	%f_{p,n}}{m_{p,n}} 
	= 
	 \sum_{q=u,d,s} f_{Tq}^{(p,n)} {\cal C}_{q} 
	+\frac{2}{27} f_{TG}^{(p,n)} 
	  \sum_{c,b,t} {\cal C}_{q}.
$$ 
	The spin-dependent part of the elastic $\chi$-nucleon 
	cross section can be written as
$$ 
\sigma^{p,n}_{{\rm SD}}(0) 
            = 4 \frac{\mu^2}{\pi}\ 3 \
	       \Bigl[ {\cal A}_{p,n} \Bigr]^2
\qquad {\rm where} \qquad 
{\cal A}_{p,n} = \sum_{u,d,s} {\cal A}_q \Delta q^{(p,n)},
\quad {\rm and} \quad
\mu = \frac{m_\chi M_{p,n}}{m_\chi+ M_{p,n}}.
\label{a}
$$
	The effective couplings ${\cal A}_{q}$ and ${\cal C}_{q}$ 
	of the neutralino-quark Lagrangian
$$
L_{\rm eff} =  
	{\cal A}_{q}
	\cdot{\bar\chi}\gamma_\mu\gamma_5{\chi}\cdot
                         \bar q  \gamma^\mu\gamma_5 q 
       +{\cal C}_{q} %\frac{m_{q}}{M_{W}}
	{\bar\chi}{\chi}\cdot \bar q q
       +O\left({1}/{m_{\tilde q}^4}\right)
$$
	which enter the cross sections one can find in  
\cite{h9401262}.
	The parameters 
	$f_{Tq}^{(p,n)}$ and $ f_{TG}^{(p,n)}$ are defined by
$$
m_p f_{Tq}^{(p)} \equiv \langle p | m_{q} \bar{q} q | p \rangle, 
%\equiv m_q B_q, 
\qquad	
f_{TG}^{(p,n)} = 1 - \sum_{q=u,d,s} f_{Tq}^{(p,n)}.
$$
	Following 
\cite{h0001005} we use the updated parameters: 
\begin{eqnarray}
\label{Proton-update}
f_{Tu}^{(p)} = 0.020 \pm 0.004, \qquad f_{Td}^{(p)} = 0.026 \pm 0.005,
\qquad f_{Ts}^{(p)} = 0.118 \pm 0.062;
\\
\label{Neutron-update}
f_{Tu}^{(n)} = 0.014 \pm 0.003, \qquad f_{Td}^{(n)} = 0.036 \pm 0.008,
\qquad f_{Ts}^{(n)} = 0.118 \pm 0.062.
\end{eqnarray}
	Our estimations of the effect of the inaccuracy 
	in the determination of $f_{Ts}^{}$ on the total event rate
	agree with those obtained before in 
\cite{h9401262} and in 
\cite{h0008336,h0004043,h0005234,h0007113}.
	For a different determination using an analytic analysis see 
\cite{h0003186}. 
        The two corridors do ovelap. 
        The inaccuracy maximally changes the proton-neutralino cross 
	section (event rate) within about one order of magnitude.
        The value chosen in this work gives probably a more pessimistic 
        view of the cross sections.
	The inaccuracy of other parameters has
	a smaller effect on the cross sections.
  
	The factors $\Delta_{i}^{(p,n)}$ parameterize the quark 
	spin content of the nucleon. 
	A global QCD analysis for the $g_1$ structure functions
\cite{Mallot}, including ${\cal O}(\alpha_s^3)$ corrections,
	corresponds to the values
\cite{h0001005}
\begin{equation}
\label{Spin-update}
\Delta_{u}^{(p)} = \Delta_{d}^{(n)} =   0.78 \pm 0.02, 	\quad 
\Delta_{d}^{(p)} = \Delta_{u}^{(n)} =  -0.48 \pm 0.02,	\quad 
\Delta_{s}^{(p)} = \Delta_{s}^{(n)} =  -0.15 \pm 0.02.
\end{equation}

   	We calculate $\Omega_{\chi} h^2_0$  following the standard
   	procedure on the basis of the approximate formula
\cite{DreesNojiriOmega,Hagelin}. 
	We take into account all channels of the $\chi-\chi$ annihilation.
   	Since the neutralinos are mixtures of gauginos and
   	higgsinos, the annihilation can occur both, via
   	s-channel exchange of the $Z^0$ and Higgs bosons and
   	t-channel exchange of a scalar particle.
   	This constrains the parameter space
\cite{KaKoRozWells,DreesNojiriOmega}.
	As mentioned in the introduction we require
	$0.1 < \Omega_\chi h^2_0< 0.3$,
	for comparison we also present our results when 
	$0.025 < \Omega_\chi h^2_0< 1$.

        Another stringent constraint is imposed by the 
	branching ratio of $b\to s\gamma$ decay, 
	measured by the CLEO collaboration to be 
	$1.0 \times 10^{-4} < {\rm B}(b\to s \gamma) < 4.2 \times 10^{-4}$.
        In the MSSM this flavor-changing neutral current 
        process receives contributions from $H^\pm$--$t$,
        $\tilde{\chi}^\pm$--$\tilde{t}$ and $\tilde{g}$--$\tilde{q}$ loops
        in addition to the standard model $W$--$t$ loop.
	This also restricts the SUSY parameter space
\cite{BerBorMasRi}.

	The masses of the supersymmetric particles are constrained 
	by the results from the high energy colliders.
	This imposes relevant constraints on 
	the parameter space of the MSSM.
	In 
\cite{h9908427} we used the following lower bounds 
	for the SUSY particles 
\cite{PDG}:
$ M_{\tilde{\chi}^+_{2}} \geq 65$~GeV for the light chargino,
$ M_{\tilde{\chi}^+_{1}} \geq 99$~GeV for the heavy chargino,
$ M_{\tilde{\chi}^0_{1,2,3}} \geq 45, 76, 127$~GeV for non-LSP
					neutralinos, respectively;
$ M_{\tilde{\nu}}      \geq 43$~GeV for sneutrinos,
$ M_{\tilde{e}_R}      \geq 70$~GeV for selectrons,
$ M_{\tilde{q}}       \geq 210$~GeV  for squarks,
$ M_{\tilde{t}_1}      \geq 85$~GeV  for light top-squark,   
$ M_{H^0}              \geq 79$~GeV  for neutral Higgs bosons,
$ M_{\rm CH}           \geq 70$~GeV  for the charged Higgs boson.
	On the basis of last LEP results 
\cite{newLEP}	
	we use now new lower limits for charginos: 
	$M_{\tilde\chi^\pm_{1,2}} \ge 100\,$GeV, 
	and neutral Higgs bosons: $m^{}_{H^0} > 100\,$GeV.

	As previously
\cite{h9908427} we explore the MSSM parameter space at the weak scale
	relaxing completely constraints following from 
	any unification assumption.
	Nevertheless we respect other available restrictions from cosmology, 
	accelerator SUSY searches, rare FCNC $b\to s\gamma$ decay, etc
\cite{h9801246,h9701301,KaKoRozWells}.

	The MSSM parameter space is determined by entries of the mass 
	matrices of neutralinos, charginos, Higgs bosons, 
	sleptons and squarks. 
	The relevant definitions one can find in 
\cite{h9908427}.
	The list of free parameters includes: 
	$\tan\beta$ is the ratio
	of neutral Higgs boson vacuum expectation values, 
	$\mu$ is the bilinear Higgs parameter of the superpotential,
	$M_{1,2}$ are soft gaugino masses, 
	$M_A$ is the CP-odd Higgs mass, 
	$m^2_{\widetilde Q}$, $m^2_{\widetilde U}$, $m^2_{\widetilde D}$ 
	($m^2_{\widetilde L}$, $m^2_{\widetilde E}$) are 
	squared squark (slepton) 
	mass parameters for the 1st and 2nd generation,        
	$m^2_{\widetilde Q_3}$, $m^2_{\widetilde T}$, $m^2_{\widetilde B}$ 
	($m^2_{\widetilde L_3}$, $m^2_{\widetilde \tau}$) 	
	are squared squark (slepton) mass parameters 
	for 3rd generation 
	and $A_t$, $A_b$, $A_\tau$ are soft trilinear 
	couplings for the 3rd generation.
	In our numerical analysis the parameters of the MSSM are 
	randomly varied in the following intervals: 
\begin{eqnarray}
\nonumber
&-1{\rm ~TeV} < M_1 < 1{\rm ~TeV}, \quad
-2{\rm ~TeV} < M_2, \mu, A_t < 2{\rm ~TeV},&
\\ \label{Scan}
&1 < \tan\beta < 50, \quad
60{\rm ~GeV} < M_A < 1000{\rm ~GeV},&\\
\nonumber
&10{\rm ~GeV}^2 < m^2_{Q_{}}, 
m^2_{L}, m^2_{Q_3}, m^2_{L_3}<10^6{\rm ~GeV}^2.&
\end{eqnarray}
	Following 
\cite{h0007202,h0008336,h0005234} 
	we assume that squarks are basically degenerate. 
	Bounds on flavor-changing neutral currents 
	imply that squarks with equal gauge
	quantum numbers must be close in mass. 
	With the possible exception of third generation squarks
	the assumed degeneracy therefore holds almost model-independently
\cite{h0007202}. 
	Therefore for other sfermion mass parameters we used the relations 
	$m^2_{\widetilde U_{}} = m^2_{\widetilde D_{}} 
	= m^2_{\widetilde Q_{}}$, 
	$m^2_{\widetilde E_{}} = m^2_{L}$, 
	$m^2_{\widetilde T} = m^2_{\widetilde B} = m^2_{Q_3}$,  
	$m^2_{\widetilde E_{3}} = m^2_{L_3}$.
	The parameters $A_b$ and $A_\tau$ are fixed to be zero.
	We consider the domain of the MSSM parameter space, 
	in which we perform our scans, as quite spread and natural. 
	Any extra expansion of it like, for example, using
	$-10$~TeV$< M_2< $~10~TeV, etc, of course, can be possible, 
	but should be considered, contrary to 
\cite{h0004043,h9908309,a0008115}, as quite unnatural.

%%%%%%%%%%%%%%%%%%%%%%%%%%%%%%%%
\section{Results and discussion}
%%%%%%%%%%%%%%%%%%%%%%%%%%%%%%%%
\subsection{Coannihilation}
	The effects of coannihilation may become important when the 
	next to the lowest supersymmetric particle (NLSP) has a mass
	which lies close to the LSP mass
\cite{yamaguchi0}. 
	The size of the effects is
	exponentially damped by the factor $e^{-\Delta_ix}$ where
	$\Delta_i=(m_i/m_{\chi}-1)$, $x=m_{\chi}/kT$ and where $m_{\chi}$ 
	is the LSP mass. 
	Because of this damping the coannihilation effects are typically
	important only for regions of the parameter space where
	the constraint $\Delta_i< 0.1$ is satisfied. 
 	Some of the possible candidates for NLSP are the light stau 
	$\tilde \tau_1$, $\tilde e_R$, the next to the lightest neutralino 
	$\chi^0_2$, and the light chargino  $\chi_1^+$. 
	It was found that in mSUGRA the upper limit on the neutralino mass 
	consistent with the current experimental constraints on the relic 
	density is extended from 200 GeV to 600 GeV
\cite{EFOSi} when the effects of 
	$\chi -\tilde\tau$ coannihilation are included.

	By means of excluding points which can give non-negligible 
	contribution to relic neutralino annihilation via
	coannihilations with other SUSY particles
	we simply have estimates of the influence of the coannihilation
	to our previous results.
	We used the constraint:
	$(m^{}_{i}-m^{}_{\rm LSP})/(m^{}_{\rm LSP}) < 0.2$, 
	where $i$ runs over next-to-LSP neutralino, 
	charginos, staus, stops etc.
	In accordance with previous estimates of 
\cite{h0001019,h0005234,h0006266,h0009065} we found that the coannihilation
	does not significantly change our main results.
	In fact, less than 20\% of the models were denied by this 
	coannihilation constraint, which in the case of 
	$0.025 < \Omega_\chi h^2_0<1$ excludes points with simultaneously 
	small $|\mu|$ ($|\mu|<500\,$GeV) 
	and large $|M_1|$ ($|M_1|>600\,$GeV), 
	allowing for a substantial non-gaugino fraction 
	of the LSP only in the region of relatively small $|\mu|$.
	If the relic abundance of neutralinos is located in the range
	$0.1 < \Omega_\chi h^2_0<0.3$, 	
	the coannihilation constraint appears less restrictive.

%%%%%%%%%%%%%%%%%%%%%%%%%%%%
\subsection{Cross sections}
\begin{figure}[h] %%%%%%%%%%%%%%%%%%%%%%%%%%%%%%%%%%%%%
\begin{picture}(100,140)
\put(-5,-44){\includegraphics{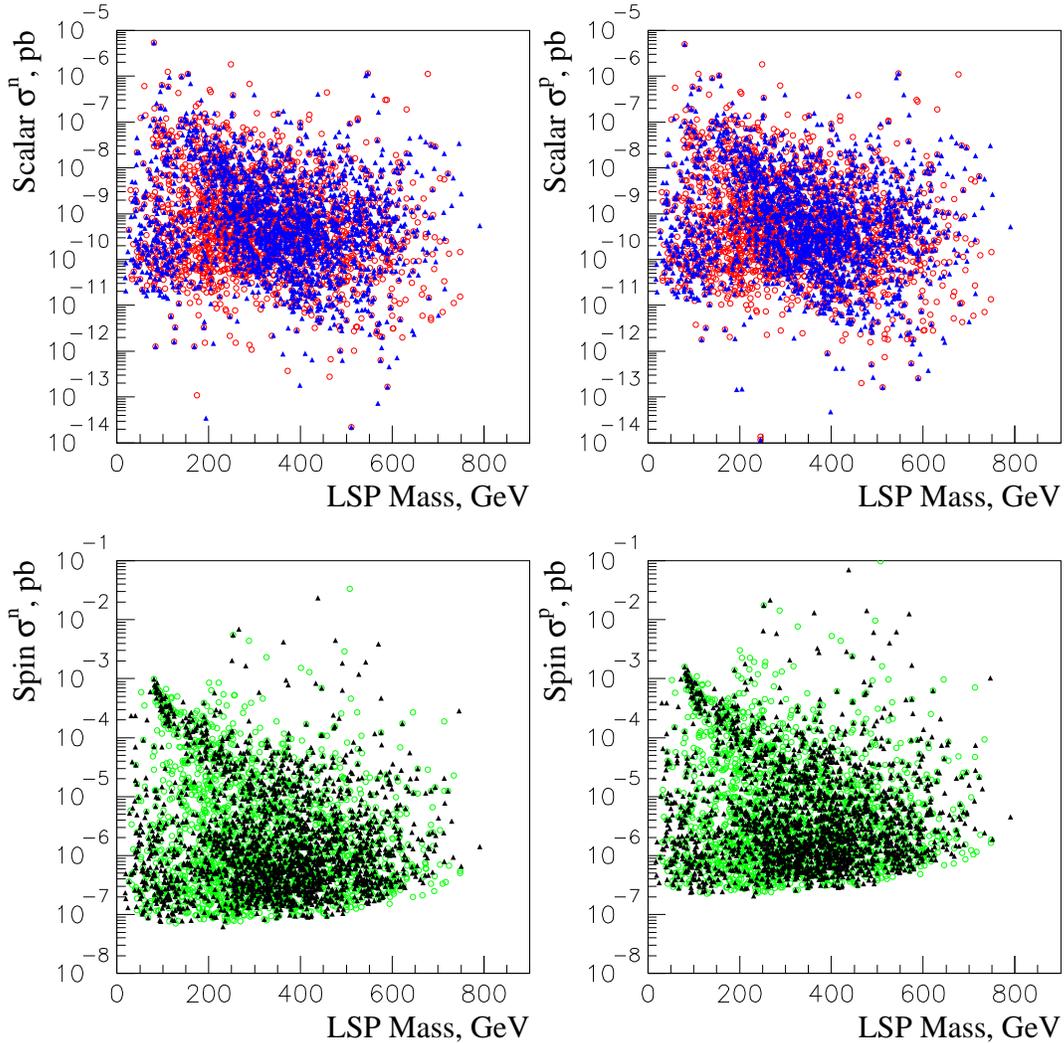}}
\end{picture}
\caption{Cross sections of spin-dependent and spin-independent
	interactions of WIMPs with proton and neutron.
	Filled triangles (light circles)
	correspond to relic neutralino density 
	$0.1 < \Omega_\chi h^2_0<0.3$ 
	($0.025 < \Omega_\chi h^2_0<1$).
\label{CrossSections}}
\end{figure}

\begin{figure}[h!] %%%%%%%%%%%%%% %angle=270 
\begin{picture}(100,145)
\put(2,-35){\includegraphics{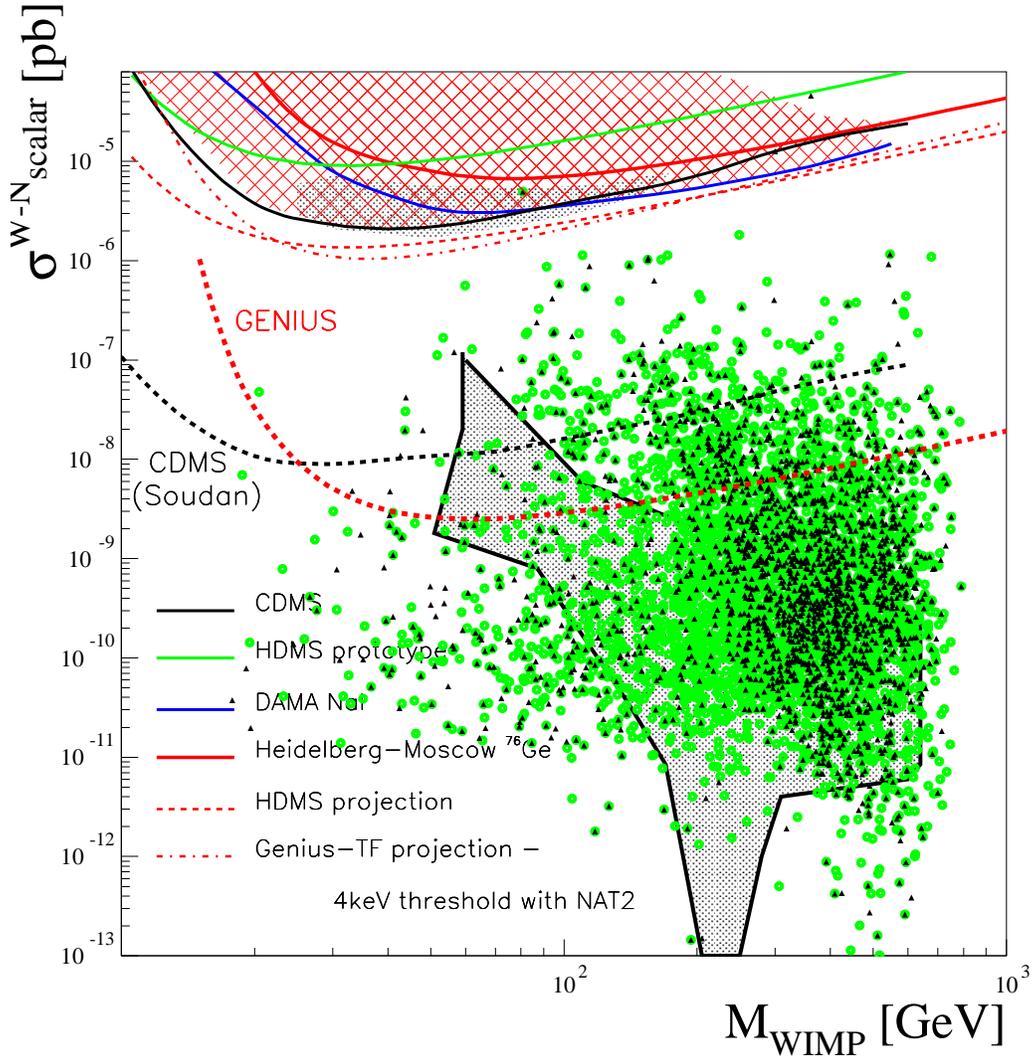}}
\end{picture}
\caption{WIMP-nucleon cross section limits in pb for scalar
	interactions as function of the WIMP mass in GeV.
	Filled circles present our calculations with updated nucleon structure 
\protect\cite{h0001005} in ``non-accelerating universe'' 
	with $0.025 < \Omega_\chi h^2_0<1$.	
	Filled triangles give the same cross section
	but when $0.1 < \Omega_\chi h^2_0<0.3$.  
	The contours (shaded area enclosed with solid curve) 
	for allowed scalar WIMP-proton cross sections from 
\protect\cite{h0001005,h0007113} are also given together with some current 
	(DAMA 
\protect\cite{DAMA}, HEIDELBERG-MOSCOW 
\protect\cite{HM-DM98}, CDMS 
\protect\cite{CDMS} and HDMS prototype
\protect\cite{HDMSpr}) 
	and future experimental exclusion curves (HDMS projection 
\protect\cite{HDMSpr}, GENIUS-TF 
\protect\cite{GENIUS-TF}, GENIUS 
\protect\cite{GENIUS} and CDMS 
\protect\cite{CDMS}).
\label{WIMP-proton}}
\end{figure}

\begin{figure}[h!] %%%%%%%%%%%%%% %angle=270 %%%%%%%%%%%%%%%%%%%%%%%
\begin{picture}(100,148)
\put(2,-35){\includegraphics{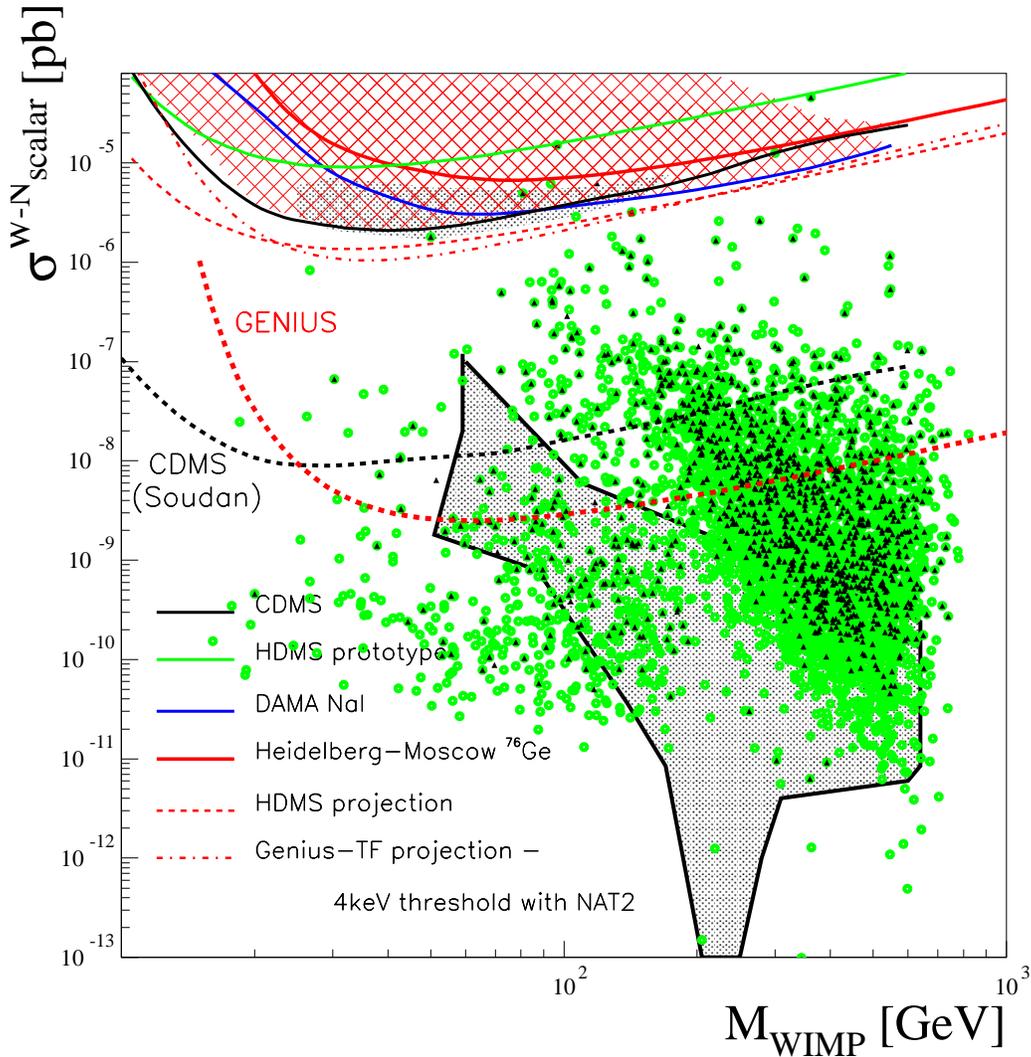}}
\end{picture}
\caption{The same as in 
fig.~\protect\ref{WIMP-proton}, but filled circles give cross sections
	for $0.1 < \Omega_\chi h^2_0<0.3$ and $\tan\beta > 20$.
	Filled triangles give the same cross section, 
	but when $\tan\beta > 40$.
	The contour from (obtained for $\tan\beta < 10$) 
\protect\cite{h0001005,h0007113} also given.
\label{WIMP-proton-tanb}}
\end{figure}

\begin{figure}[th!] %%%%%%%%%%%%%%%%%%%%%%%%%%%%%%%%%%%%%
\begin{picture}(100,150)
\put(-3,-37){\includegraphics{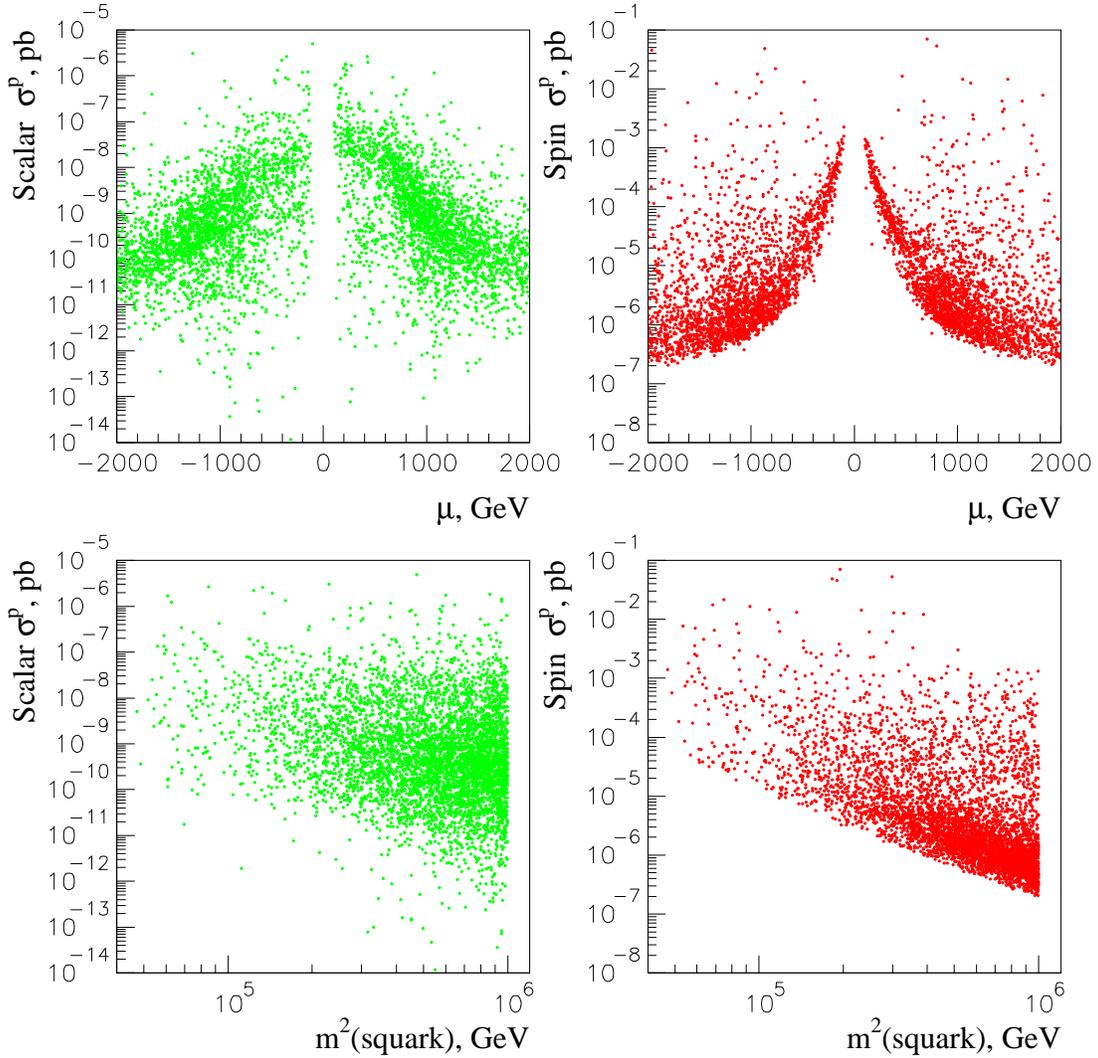}}
\end{picture}
\caption{Cross sections of of WIMP-proton 
	spin-dependent and spin-independent
	interactions as function of 
	input parameters $\mu$ (upper panel) and 
	$m^2_{Q_{}}$ (lower panel) 
	obtained with $0.1 < \Omega_\chi h^2_0<0.3$.
\label{CrossSections-E}}
\end{figure}

\begin{figure}[th!] %%%%%%%%%%%%%%%%%%%%%%%%%%%%%%%%%%%%%
\begin{picture}(100,150)
\put(-5,-37){\includegraphics{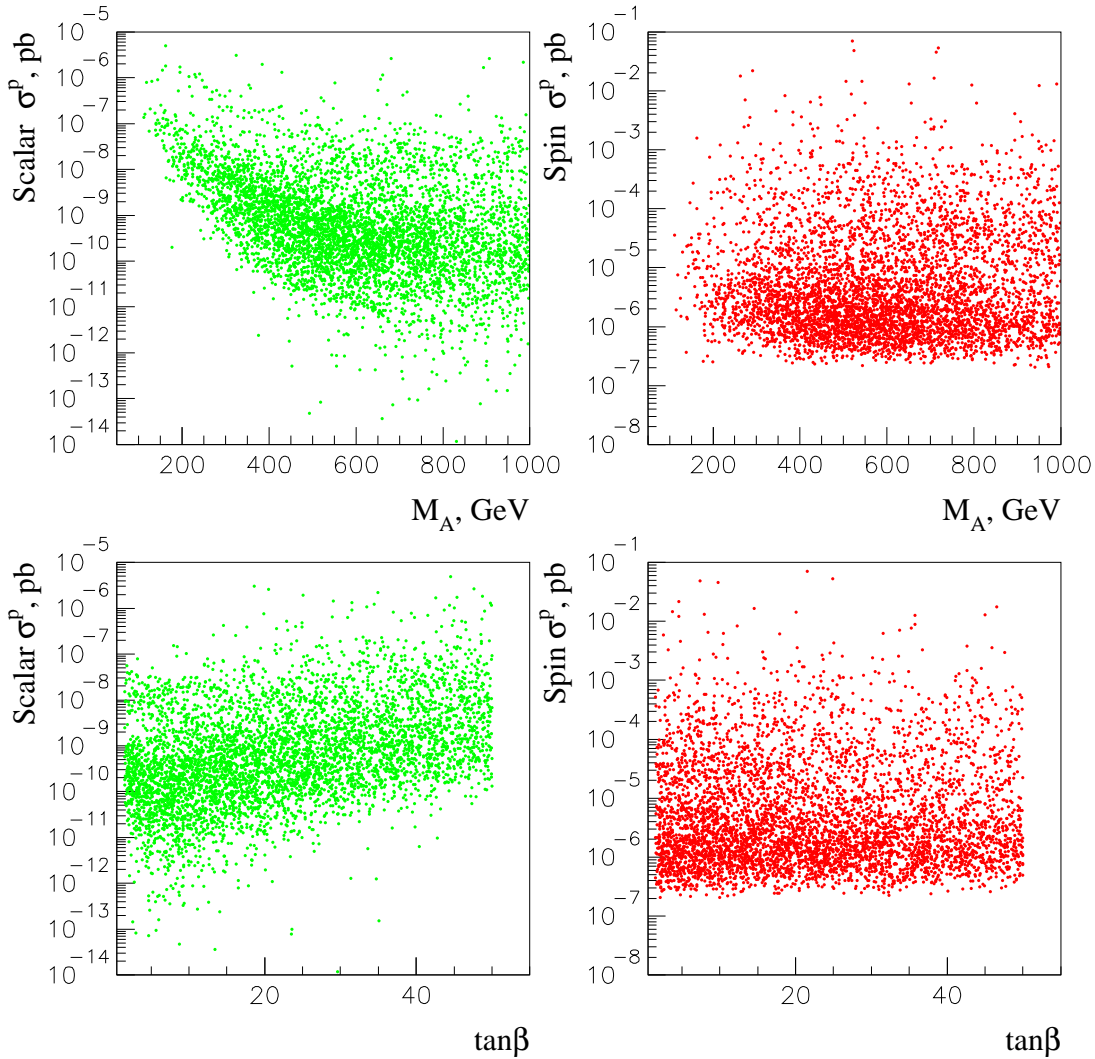}}
\end{picture}
\caption{Cross sections of WIMP-proton 
	spin-dependent and spin-independent
	interactions as function of 
	input parameters $M_A$ (upper panel) and 
	$\tan\beta$ (lower panel) 
	obtained with $0.1 < \Omega_\chi h^2_0<0.3$.
\label{CrossSections-D}}
\end{figure}

%%%%%%%%%%%%%%%%%%%%%%%%%%%%
	Our calculations with the updated nucleon structure 
\cite{h0001005} for the WIMP-nucleon cross section of both spin and scalar
	interactions as function of the WIMP mass are depicted 
	below as scatter plots 
(Figs.~\ref{CrossSections}--\ref{SpinandScalar}).
	
	The use of the updated parameters
(\ref{Proton-update})--(\ref{Spin-update}) does not change significantly
	the general distribution of points over the scatter plots
	as compared with calculations with earlier nucleon parameters 
\cite{Cheng,Gasser,EMC} used in  
\cite{h9908427,h9606261,h9608241,h9401262}.

 %%%%%%%%%%%%%%%%%%%%

	Scatter plots with individual cross sections of spin-dependent 
	and spin-independent interactions of WIMPs with proton and neutron
	are given in 
fig.~\ref{CrossSections} as functions of the LSP mass.
	In the figure light circles correspond to cross sections
	calculated under the old assumption that  
	$0.025 < \Omega_\chi h^2_0<1$.
	Filled triangles give the same cross section but the constraint 
	on the flat and accelerating universe is imposed by 
	$0.1 < \Omega_\chi h^2_0<0.3$.
	One can see that the reduction of the allowed domain
	for the relic density does not significantly affect
	spin-dependent and the spin-independent WIMP-nucleon cross sections, 
	i.e. restriction to a flat and accelerating universe
	weakly affects these cross sections.

	The different behavior of these cross sections with 
	mass of the LSP can be seen from the plots.
	There is a more stringent 
	lower bound for the spin-dependent cross section.
	It is at a level of $10^{-7}\,$pb, which is 
	about an order of magnitude larger then the one presented in 
\cite{h0007113}, where for small $\tan\beta$
	($\tan \beta = 3,~\mu<0$, and small $m_\chi$)
	the effect of a cancellation induced by the difference in signs 
	between $\Delta_u$ and $\Delta_{d,s}$
(\ref{Spin-update}) was reported.
	Aside from the cancellation, the spin-dependent cross section peaks 
	at about $10^{-4}$ pb and drops rapidly as $m_\chi$ increases 
	down to $10^{-7\div8}$~pb at $m_\chi \approx 600$~GeV
\cite{h0007113}.
	We have checked that special consideration 
	of the low $\tan\beta$ regime supplies us also with 
	smaller cross section values for spin-dependent
	interactions, which do not enter in contradictions with
\cite{h0007113}.	

	Such a cancellation was found also in scalar cross sections
	for $\tan \beta = 10$ and $\mu < 0$.
\cite{h0007113}. 
	In this case Higgs exchange is dominant.
	The cancellation in the mSUGRA is due to the cancellation
	between the up-type contribution (which is negative) and the
	down-type contribution, which is initially positive but decreasing, 
	eventually becoming negative as we increase $m_{\chi}$.

%%%%%%%%%%%%%%%%%%%%

	If 
fig.~\ref{WIMP-proton} filled circles present our calculations 
	when constraints due to an accelerating universe 
	are not applied and as in 
\cite{h9908427,h0008022} we hold $0.025 < \Omega_\chi h^2_0<1$.	
	Filled triangles give the same cross section,
	but using as 
\cite{h0001005,h0001019,h0004043,h0005234,h0007113,h0006266,h0009065,h0010203}
	the boundary  $0.1 < \Omega_\chi h^2_0<0.3$.  
	The contours for allowed scalar WIMP-proton 
	cross sections from 
\cite{h0001005,h0007113} are also given together with some current (DAMA 
\cite{DAMA}, CDMS 
\cite{CDMS}, HEIDELBERG-MOSCOW 
\cite{HM-DM98}, HDMS prototype 
\cite{HDMSpr})
        and future-expected experimental 
	exclusion curves 
(HDMS \cite{HDMSpr}, GENIUS-TF \cite{GENIUS-TF}, GENIUS \cite{GENIUS} and 
CDMS \cite{CDMS}).
	This figure allows one to see the influence of the 
	flat and accelerating universe on the distribution
	of WIMP-proton scalar cross section.
	The reduction left only 25\% of points but
	nevertheless the distribution of the remaining points
	differs only slightly from the one obtained with
	$0.025 < \Omega_\chi h^2_0<1$.	
	The models with very small cross sections as well as 
	models with very large cross sections (in fact 
	experimentally excluded) still persist. 

%%%%%%%%%%%%%%%%%%%%

	One also can conclude that there is no
	contradiction between the result of
\cite{h0001005,h0007113} 
	obtained in the mSUGRA with minimal number 
	of free parameters 
	and our phenomenological scan, which also allows models
	with very small cross sections. 

	While we have $1 < \tan\beta < 50$, the contour from 
\cite{h0001005,h0007113} was
	obtained under the assumption that $\tan\beta \le 10$
	to avoid some uncertainties in the treatment of 
	radiative corrections in the renormalization-group evolution
	of the MSSM parameters which affect the relic density
	calculations
\cite{h0007113}. 
	As noticed by many groups
\cite{h9401262,h9706509,h9908427,h0001019,h0006266,h0009065,h0010203},
	the scalar cross section of elastic WIMP-nucleon
	scattering increases with $\tan\beta$.
	As can be seen from 
fig.~1 of 
\cite{h9908427} $\tan\beta$ 
	seems to be 
	the only SUSY parameter 
	with which the lower bound
	of the direct detection rate has the tendency to increase.
	The majority of the points at the scatter plots in fig.  
\ref{WIMP-proton-tanb} 
	are shifted to the domain of larger cross section 
	with increase of $\tan\beta$.

	In general the increase of 
	$\tan\beta$ effectively relaxes the $\mu$ constraint in mSUGRA 
	(it allows $\mu$ to be smaller) and results 
	in a non-negligible Higgsino component followed by 
	significantly larger scalar cross section (see for example
\cite{h9908427}).

	We also respect as before 
\cite{h9401262,h9608241,h9908427}
	non-universality of soft supersymmetry-breaking
	masses in the scalar and the gaugino sectors (see
	list of free parameters in  
(\ref{Scan})),
	resulting in larger cross sections, as noted in   
\cite{h0005234,h0007113}.

	The spin-dependent and spin-independent WIMP-proton
	cross sections as functions of input parameters
	$\mu$, $m^2_{Q}$, $M_A$, $\tan\beta$ %(see (\ref{Scan})) 
	are depicted in figs.
\ref{CrossSections-E} and
\ref{CrossSections-D}. 
 %%%%%%%%%%%%%%%%%%%%%%%%%%%%%%%%%%%%%%%%%

	There is no noticeable dependence of these scatter plots 
	on the other free parameters from  	
(\ref{Scan}), for which we therefore do not show scatter plots. 
	One can see from 
fig.~\ref{CrossSections-E} the similarity of the scatter plots  
	for spin-dependent and and scalar cross sections as functions of  
	$\mu$ and $m^2_{Q_{}}$. 
	Decrease of both lower bounds of the cross sections 
	with $m^2_{Q_{}}$ occur due to increase
	of masses of squarks, which enter the s-channel intermediate states.
	The only visible difference concerns more sharp lower bounds for 
	the spin-dependent cross section.
	Both spin-dependent and spin-independent
	cross sections increase when $|\mu|$ decreases, in agreement with 
\cite{h9908427,h0004043,h0008022,h0001019}.
	It is not easy to trace the tendency in mSUGRA
	models because the parameter $\mu$ there is 
	strongly constrained by the electroweak symmetry breaking
	condition (see, for example, 
\cite{h0007113}).

	The increase of the scalar cross sections generally is connected 
	with an increase of the Higgsino admixture of the LSP and increase 
	of Higgsino-gaugino interference which enters this cross section
\cite{h0004043,h0006266,h0001019}.
	The reason of the Higgsino growth can be 
	non-universality of scalar soft masses
\cite{h0001019}, variation of intermediate unification scale
\cite{h0006266}, or new focus point regime of supersymmetry 
\cite{h0004043}.
 %%%%%%%%%%%%%%%%%%%%

	For example, as given in 
\cite{h0006266},
	the smaller the intermediate scale $M_I$ is, the larger 
	the Higgsino components become. 
	In particular, for $M_I=10^{16}$ GeV the LSP is mainly Bino,
	the Higgs-neutralino-neutralino couplings are suppressed
	and therefore the cross sections are small.
	However for $M_I=10^{11}$ GeV the Higgsino contributions 
	become important and even dominant
	with the consequence of larger cross sections.
	It is also worth noting  that, for any fixed value of $M_I$, 
	the larger $\tan\beta$ is, 
	the larger the Higgsino contributions become.

	Also it is claimed 
\cite{h0004043} that in the specific context of minimal supergravity
	(focus point regime), 
	a cosmologically stable mixed gaugino-Higgsino state 
	emerges as an excellent, robust dark matter candidate. 
	The claim relies on recent arguments,
	that all squark and slepton masses can be
	taken well above 1 TeV with no loss of naturalness on the basis of 
	a seemingly reasonable objective definition of naturalness
\cite{h9908309}. 
	The mathematical basis of this result
	is the existence of focus points in renormalization group
	trajectories, which render the weak scale (i.e., the Higgs
	potential) largely insensitive to variations in unknown supersymmetry
	parameters%
\footnote{The background of the approach can be questioned
\cite{Kazakov} due to a possibility to shift the focus point for the Higgs 
	mass parameter right to the GUT scale by means ofcappropriate choice 
	of the initial condition for the top Yukawa coupling
\cite{h9902290}. 	
	Anyway from a phenomenological point of view the approach is 
	interesting.}.  
	While in these models the squark and slepton masses are 
	unusually large, the electroweak gaugino and
	Higgsino particle masses are generically well below 1 TeV.
	The increase of the common soft scalar mass $m_0$
	far beyond 1 TeV associated with decrease of $|\mu|$ below the 
	gaugino masses, $M_1, M_2$, leads to significant mixing between 
	Higgsino and gaugino states accompanied by Higgs boson 
	diagrams enhancement. 
	A net result are again large scalar cross sections.
	
	The arguments presented above do not work in SUGRA
\cite{h0007113}.
	The LSP as Higgsino-like is almost excluded by LEP constraints
\cite{EFGOS} even if the assumptions of universal soft supersymmetry 
	breaking are relaxed, and Higgsino dark matter is certainly excluded 
	if universality is assumed, as is the case here. 
	In addition to the LEP constraints, this is because
	the value of $\mu$ is predicted as a function of 
	$m_{1/2}$ and $m_0$, placing the LSP firmly 
	in the Bino-like region. 
	The same considerations exclude 
	an LSP with mixed Higgsino/gaugino content.

	In the SUGRA framework of 
\cite{h0007113}
	the elastic scattering cross sections, which are predicted 
	for the LSP mass $m_\chi$ lie in a comparatively narrow band. 
	This is essentially because the LSP is always mainly
	Bino-like, so its couplings do not depend
	greatly on other MSSM parameters such as $m_0$. 
	The principal causes of broadening are the uncertainties 
	in the hadronic inputs and the possibilities of cancellations 
	that may reduce the cross sections for
	some specific values of the constrained MSSM parameters
\cite{h0007113}.

\smallskip

	Figure 
\ref{CrossSections-D} shows that while the spin cross section displays
	almost full insensivity to $\mu$ and $M_A$ 
	(Higgs bosons do not contribute)
	the scalar cross section possesses
	remarkable dependence on these parameters.
 	The cross section rather quickly drops with growth
	of the CP-odd Higgs mass $M_A$ and increases with 
	$\tan\beta$ in accordance with results of 
\cite{h9401262,h9706509,h9908427,h0001019,h0006266,h0009065,h0010203}.

	The different $\tan\beta$- and $M_A$-dependence 
	of spin-dependent and spin-independent cross section
	as well as general about-4-order-of-magnitude 
	excess of spin-dependent cross section over
	spin-independent
	cross sections may be important for observations
\cite{h0005041,h0010036}.		

%%%%%%%%%%%%%%%%%%%%%%%%%%%%%
\subsection{Role of the spin}
\begin{figure}[th!] %%%%%%%%%%%%%%%%%%%%%%%%%%%%%%%%%%%%%
\begin{picture}(100,140)
\put(-5,-41){\includegraphics{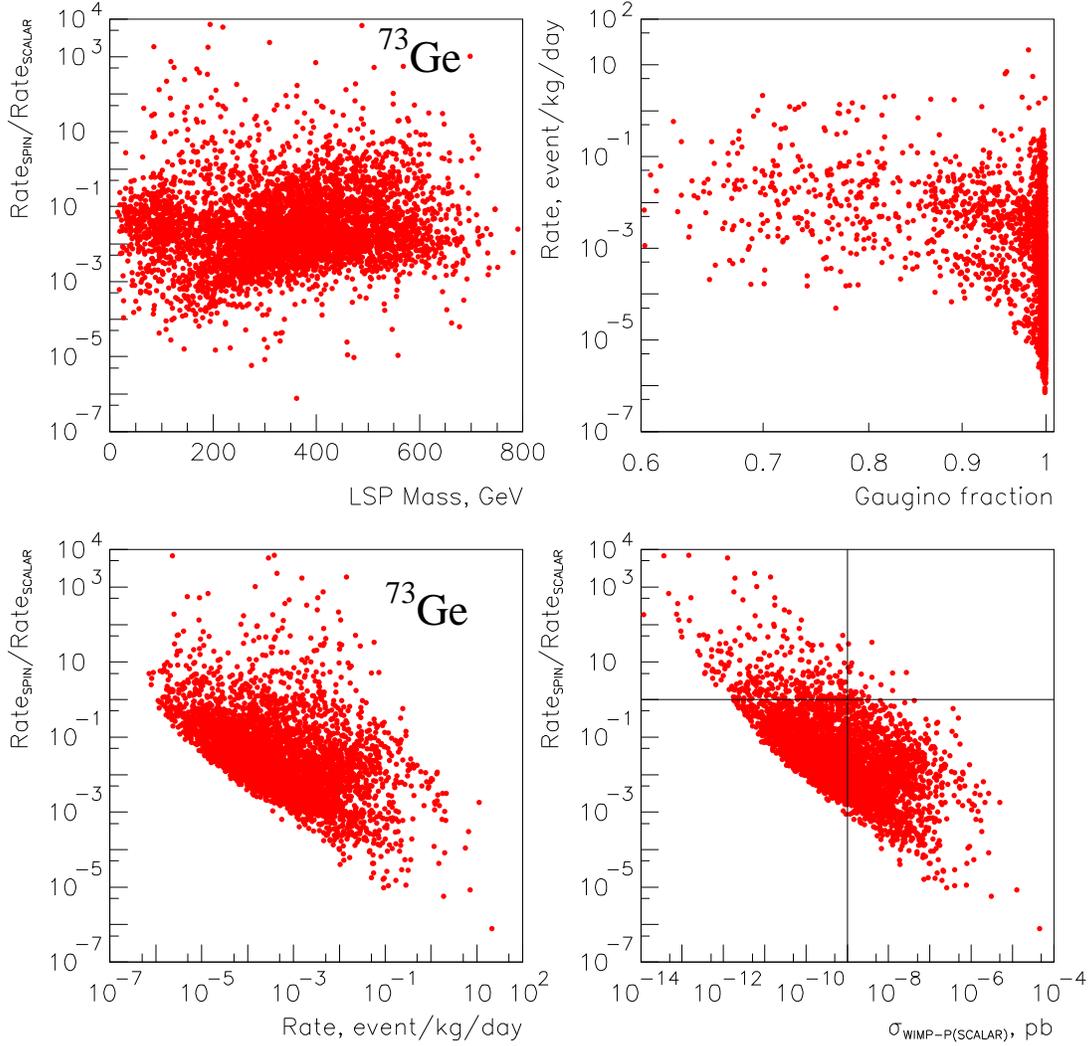}}
\end{picture}
\caption{Ratio of spin-dependent event rate to the   	
	spin-independent event rate in $^{73}$Ge isotope
	as function of LSP mass (upper left), 
	total (spin-dependent plus spin-independent) event rate (lower left) 
	and scalar cross section of neutralino-proton interaction 
	(lower right) obtained with $0.1 < \Omega_\chi h^2_0<0.3$.
	The vertical line gives the expected sensitivity of GENIUS
\protect\cite{GENIUS}. 
	In the region above the horizontal line the spin contribution 
	dominates.
	The total event rate versus 
	gaugino fraction of LSP $P$ also given (upper right), 
\label{SpinandScalar}}
\vspace*{-0.5\baselineskip}
\end{figure}

%%%%%%%%%%%%%%%%%%%%%%%%%%%%%
	To be more definite with the statement claimed above, in 
fig.~\ref{SpinandScalar} we present a comparison of total spin-dependent 
	versus total spin-independent event rates in $^{73}$Ge 
	(spin$\,=9/2$)~--- 
	as representative and one of the most promising isotopes 
	for future construction of high-sensitivity detectors.

	Figure 
\ref{SpinandScalar}
	shows the weak dependence of the ratio on mass of the LSP
	with the mean value being approximately 0.01--0.1.
	There are very large and very small values for the
	ratio practically for any given mass of the LSP.
	The spin-independent (scalar) contribution
	obviously dominates in the domain of  
	large expected rates in the Germanium detector
	($R>0.1\,$events$/$day$/$kg) 
	as was obtained before (see, for example 
\cite{h9401262}). 
	But as soon as the total rate drops down to
	$R<0.01\,$events$/$day$/$kg or, equivalently, the
	scalar neutralino-proton cross section becomes
	smaller than $10^{-9}\div10^{-10}\,$pb,
	the spin-dependent interaction may produce
	a rather non-negligible contribution to the total event rate.
	Moreover if the scalar cross section decreases further
	($\sigma < 10^{-12}\,$pb), it becomes obvious that 
	the spin contribution alone saturates 
	the total rate and protects it (see lower bounds in fig.
\ref{CrossSections-E} and 
\ref{CrossSections-D}) from decreasing below
	$R\approx 10^{-6}\div 10^{-7}\,$events$/$day$/$kg
\cite{h9908427}.

	This observation could be quite important for experiments actually 
	looking for direct {\em detection}\ of dark matter, but 
	not only for exclusion plots. 
	Indeed, while scalar cross sections governed mostly by Higgs exchange 
	can be rather small (when Higgs masses remain large enough,
	for example in the Next-to-Minimal Supersymmetric Standard Model
\cite{h9802344}) the spin cross section can not be arbitrary small, 
	because the mass of the $Z$ boson, which gives the dominant 
	contribution, is well defined, provided one ignores any 
	possible fine-tuning cancellations 
\cite{h0007113}.

	Therefore, if an experiment with sensitivity 
	$10^{-5}\div 10^{-6}\,$events$/$day$/$kg
	fails to detect a dark matter signal,
	an experiment with higher sensitivity (and non-zero spin target) 
	will be able to detect dark matter particles 
	only due to the spin neutralino-quark interaction.

 %%%%%%%%%%%%%%%%%%%%

%%%%%%%%%%%%%%%%%%%%
\section{Conclusion}
%%%%%%%%%%%%%%%%%%%%
\vspace*{-0.5\baselineskip}
	Recent measurements in modern cosmology
	have changed the expected fraction of the 
	cold dark matter in the universe, 
	new results for the nucleon structure were obtained,
%	the new theoretical approaches to 
%	detection of the dark matter appeared,
	a new generation of high-sensitivity
	experimental detectors are under consideration.
	All these changes stimulated us to recalculate 
	our previous analysis concerning detection 
	of cold dark matter. 

	To this end 
	we explored the MSSM parameter space at the weak scale 
	where new accelerator and 
	cosmological constraints are respected.
 	We restrict the relic neutralino density to be in the range 
	$0.1\le \Omega_{\chi} h^2 \le 0.3$.
	We considered the 
	variation of the spin-independent
	and spin-dependent WIMP-nucleon cross sections 
	and of the expected event rate in $^{73}$Ge,  
	with parameters of the MSSM,  
	uncertainties of the nucleon structure
	and other theoretical assumptions like 
	universality and coannihilation. 

	The main results of the exploration can be summarized
	as follows.

1. 	The results of our updated calculations 
	fall in general agreement with calculations 
	performed in mSUGRA as well as with 
	other less restrictive approaches, which allowed
	larger variation of the cross sections and detection rates.

2.	The use of the updated parameters of the nucleon structure
	does not change significantly the 
	general distribution of points over the scatter plots
	as compared with calculations with earlier nucleon parameters. 

3.	In accordance with previous estimations 
	we found that the coannihilation
	does not significantly change our main conclusions.
	We understand that our estimation of the
	coannihilation effect is somewhat indirect, but  
	in the effMSSM approach there is no stringent 
	correlation between parameters, which sometimes
	makes the coannihilation channels inevitable.
	
4.	The new cosmological constraint on the relic neutralino
	density (due to flat and accelerating universe)
	which is numerically used in the form 
	$0.1 < \Omega_\chi h^2_0<0.3$ in our approach does not 
 	significantly affect the resulting scatter plots
	for neutralino-nucleon elastic cross sections.

5.	To single out (mostly in the mSUGRA) 
	theoretically a dominant contribution 
	to the cross section or event rate 
	one usually relies on the 
	knowledge of the LSP composition.
	For example, as discussed through this paper, 
	if the Bino fraction is large, then the 
	cross section is small.
 	Numerically the situation is less transparent.
	As seen  from fig.
\ref{SpinandScalar} (upper right) the 
	overwhelming majority of points 
	(the region of highest point density) has $P \approx 1$, or
	very small Higgsino admixture and 
	one should expect negligible event rate.
	Nevertheless this is not the case.
	There are a lot of points with sizable event rate
	for $P \approx 1$.
	Therefore qualitative estimations of the 
	dominance of the given contribution on the basis, for example,  
	of large gaugino fraction of the LSP ($P>0.9$) 
	can be quantitatively not always correct.  

%	Therefore qualitative estimations of the cross sections 
%        on the basis of 
%	dominance of a given contribution to the LSP, for example,  
%	of large gaugino fraction of the LSP ($P>0.9$) 
%	can be quantitatively not always correct.  

6.	We notice 
	that for targets with spin-non-zero nuclei
	it might be the {\it spin-dependent interaction}\ that 
	determines the lower bound for the direct detection rate
	when the cross section of the scalar interaction 
	drops below about $10^{-12}\,$pb.
	If this occurs the spin nuclear detectors would have 
	notable advantage comparing with spinless detectors, or 
	may become the only way to observe SUSY via direct 
	dark matter detection.

	Finally we would like to stress again the fact, clearly seen
	from figs.
\ref{WIMP-proton} and 
\ref{WIMP-proton-tanb}, that 
	to reliably investigate the 
	SUSY parameter space and therefore 
	to have a chance to beat accelerator experiments in 
	searching for (or discovery of) the new physics (supersymmetry)
	one needs a GENIUS-like detector.

\smallskip
Acknowlegment: 
	The authors would like to thank Prof. Pran Nath for useful
	discussions. 
	One of us (V.B.) thanks the Max Planck Institut fuer Kernphysik 
	for the hospitality extended to him during his stay at Heidelberg.
        The investigation was supported in part (V.B.)
        by RFBR Grant 00-02--17587. 

%%%%%%%%%%%%%%%%%%%%%%%%%%%%

%%%%%%%%%%%%%%%%%%%%
\end{document}